\documentclass[a4paper,11pt]{article}
\usepackage{pos}

\title{Flavor number dependence of QCD at finite density by the complex Langevin method}

\author*[a]{Yusuke Namekawa}
\author[b]{Yuhma Asano}
\author[c]{Yuta Ito}
\author[d,e]{Takashi Kaneko}
\author[d,e]{Hideo Matsufuru}
\author[d,e]{Jun Nishimura}
\author[f]{Asato Tsuchiya}
\author[g]{Shoichiro Tsutsui}
\author[h,i]{Takeru Yokota}

\affiliation[a]{Department of Physics, Kyoto University, Kyoto 606-8502, Japan}
\affiliation[b]{Faculty of Pure and Applied Sciences, University of Tsukuba, 1-1-1 Tennodai, Tsukuba, Ibaraki 305-8577 Japan}
\affiliation[c]{National Institute of Technology, Tokuyama College,  Gakuendai, Shunan, Yamaguchi 745-8585, Japan}
\affiliation[d]{High Energy Accelerator Research Organization (KEK), 1-1 Oho, Tsukuba, Ibaraki 305-0801, Japan}
\affiliation[e]{Department of Particle and Nuclear Physics, School of High Energy Accelerator Science, Graduate University for Advanced Studies (SOKENDAI), 1-1 Oho, Tsukuba, Ibaraki 305-0801, Japan}
\affiliation[f]{Department of Physics, Shizuoka University, 836 Ohya, Suruga-ku, Shizuoka 422-8529, Japan}
\affiliation[g]{Quantum Hadron Physics Laboratory, RIKEN Nishina Center, Wako, Saitama 351-0198, Japan}
\affiliation[h]{
Interdisciplinary Theoretical and Mathematical Sciences Program (iTHEMS), RIKEN, Wako, Saitama 351-0198, Japan}
\affiliation[i]{Institute for Solid State Physics, The University of Tokyo, Kashiwa, Chiba 277-8581, Japan}

\emailAdd{namekawa@gauge.scphys.kyoto-u.ac.jp}

\abstract{
We discuss the flavor number dependence of QCD at low temperature and high density by the complex Langevin method.
In our previous work, the complex Langevin method is confirmed to satisfy the criterion for correct convergence in certain regions, such as $\mu_{\rm q} / T = 5.2-7.2$ on $8^3 \times 16$ and $\mu_{\rm q} / T = 1.6-9.6$ on $16^3 \times 32$ using $N_{\rm f} = 4$ staggered fermion at $\beta = 5.7$.
We extend this study to more realistic flavor cases, $N_{\rm f} = 2, 2 + 1, 3$, using Wilson fermions.
We present the flavor number dependence of the validity regions of the complex Langevin method and the quark number.
}

\FullConference{
 The 38th International Symposium on Lattice Field Theory, LATTICE2021
  26th-30th July, 2021
  Zoom/Gather@Massachusetts Institute of Technology
}

\note{KEK-TH-2371, KUNS-2903, RIKEN-iTHEMS-Report-21, RIKEN-QHP-510}

\begin{document}
\maketitle

\section{Introduction}

QCD at high density attracts significant physical interest.
It is expected to have a rich phase structure, which may be realized in the neutron star.
A lot of experiments, including heavy ion collisions and observation of neutron stars, search for the physics of QCD at high density.
On the theoretical side, the exploration with the conventional Monte Carlo method is prevented by the sign problem.
Several approaches are developed and utilized for quantitative study of the finite density QCD, as overviewed in this conference~\cite{Guenther:2021lat}.

One of the promising approaches is the complex Langevin method (CLM)~\cite{Parisi:1984cs, Klauder:1983sp}.
The CLM needs no interpretation of the Boltzmann weight as a probability density.
The CLM does not suffer from the sign problem.
It is also advantageous that the simulation of the odd-flavor is straightforward.
In contrast, the CLM suffers from the wrong convergence problem.
Correctness of the CLM is guaranteed only if the justification conditions are satisfied.
It is crucial to identify the validity region of the CLM for QCD at finite density.

We have successfully identified the validity region of the CLM at finite temperature $T$ and quark chemical potential $\mu_{\rm q}$, such as $\mu_{\rm q} / T = 5.2$--$7.2$ on $8^3 \times 16$ and $\mu_{\rm q} / T = 1.6-9.6$ on $16^3 \times 32$ lattices using the four-flavor staggered quark~\cite{Ito:2020mys}.
The validity is confirmed by the criterion based on the drift histogram~\cite{Nagata:2016vkn}.
If the histogram of the magnitude of the drift term falls off exponentially or faster, the CLM guarantees to give a correct result.
If it shows a power-law tail, the CLM can converge to a wrong result.
We measured the quark number in the validity region and found that it has a plateau structure, as observed in the free theory with the naive lattice fermion~\cite{Matsuoka:1983wy} as well as the one-loop calculation on a small S$^3$ in the continuum QCD~\cite{Hands:2010zp}.

Encouraged by our previous work, we extend this study to more realistic flavor cases.
The number of flavor $N_{\rm f}$ is changed from $N_{\rm f} = 4$ required by the staggered fermion to $N_{\rm f} = 2, 2 + 1, 3$ using Wilson fermion.
We search for the validity region of low temperature and high density QCD via the CLM, varying $N_{\rm f}$, $\beta$ and the quark mass $m_{\rm q}$ as a function of $\mu_{\rm q}$.
Creating the map of the validity region is an indispensable first step for a practical calculation by use of the CLM.


\section{Formulation}
\label{sec:formulation}

We perform a simulation using the CLM in the presence of the chemical potential.
We use the plaquette gauge and Wilson fermion actions.
The gauge action $S_{g}[U]$ is defined by
\begin{equation}
 S_{g}[U] := -\frac{\beta}{6}
              \sum_{x}\sum_{\mu \neq \nu}
              \mathrm{tr} \left( U_{x,\mu} U_{x+\hat{\mu},\nu} U_{x+\hat{\nu},\mu}^{-1} U_{x,\nu}^{-1}
                          \right) \ ,
 \label{eq:gauge_action}
\end{equation}
where $\beta := 6 / g^2$, and $U_{x,\nu}\in$ SU(3) is a link variable in the $\nu = 1, 2, 3, 4$ direction with a unit vector in the $\nu$-direction $\hat{\nu}$.
The fermion action $S_{f}[U]$ for a fermion field $\psi$ is defined by
\begin{align}
 S_{f}[U]  :=& \sum_{x,y} \overline{\psi}_x M_{xy} \psi_y \ ,
 \label{eq:fermion_action}
 \\
 M_{xy}[U] :=& \delta_{xy}
              - \kappa_{q}
                 \sum_\nu \left[  (1 - \gamma_\nu) e^{ \mu_{\rm q} \delta_{\nu 4}} U_{x, \nu}      \delta_{x+\hat{\nu}, y}
                                 +(1 + \gamma_\nu) e^{-\mu_{\rm q} \delta_{\nu 4}} U_{x, \nu}^{-1} \delta_{x, y+\hat{\nu}}
                          \right] \ .
 \label{eq:dirac_op}
\end{align}
The hopping parameter $\kappa_q$ is related to the bare quark mass $m_{\rm q}$ through $m_{\rm q} = (1/2) (1 / \kappa_q - 1 / \kappa_c)$ with the critical hopping parameter $\kappa_c$.
For $N_{\rm f} = 2, 3, 4$ cases, we employ degenerate quark masses.
For $N_{\rm f} = 2 + 1$ case, we employ light and heavy quarks represented by $\kappa_{l}$ and $\kappa_{h}$, respectively.
We assign a periodic boundary condition in each direction,
except an anti-periodic boundary condition in the temporal direction for the fermion.
The determinant $\det M[U]$ becomes complex by finite $\mu_{\rm q}$.
It makes the Boltzmann weight complex and causes the sign problem.

We overcome the sign problem via the CLM.
The link variables are complexified to $\mathcal{U}_{x,\nu}\in$ SL(3,C).
They are evolved by the complex Langevin equation with the stepsize $\varepsilon_L$ using an improved second-order Runge-Kutta algorithm~\cite{Bali:2013pla},
\begin{equation}
 \mathcal{U}_{x,\nu}(t + \varepsilon_L) =
 \exp \left\{ i \left( -\varepsilon_L v_{x,\nu}[\mathcal{U}(t)] + \sqrt{\varepsilon_L} \eta_{x,\nu}(t)
                \right)
      \right\}
      \mathcal{U}_{x,\nu}(t) \ ,
 \label{eq:Langevin_eq}
\end{equation}
where $\eta_{x,\nu}(t)$ is a noise term, generated from the Gaussian distribution $\exp[- (1/4) \mathrm{tr} \, \eta_{x,\nu}^{2}(t)]$.
The noise average satisfies the orthogonal relation,
\begin{equation}
 \langle \eta_{x,\mu}^{ij}(s) \eta_{y,\nu}^{kl}(t) \rangle_{\eta}
 = 2 \delta_{x y} \delta_{\mu\nu} \delta_{st} 
   \left( \delta_{il} \delta_{jk} - \frac{1}{3} \delta_{ij} \delta_{kl}
   \right) \ .
\end{equation}
The fermion drift term $v_{x,\nu}^{(f)}$ and the gauge drift term $v_{x,\nu}^{(g)}$ are defined by
\begin{equation}
 v_{x,\nu}^{(f,g)}[U] := \left. \sum_{a=1}^8 \lambda_{a} \frac{\partial}{\partial \alpha} S_{f,g}[e^{i \alpha \lambda_{a}} U_{x,\nu}] \right|_{\alpha=0} \ .
 \label{eq:drift_term}
\end{equation}
An estimate for the observable $\mathcal{O}(\mathcal{U})$ is obtained by the average over the fictitious time $t$,
\begin{equation}
 \bar{\mathcal{O}} :=
 \frac{1}{t_{\rm total}} \int_{t_{0}}^{t_{0} + t_{\rm total}} dt\,\mathcal{O}(\mathcal{U}(t)) \ ,
 \label{eq:observable}
\end{equation}
where $t_0$ is the thermalization time, and $t_{\rm total}$ is the total complex Langevin time.
In this work, the observable is the quark number defined by
\begin{equation}
 N_{\rm q} 
 := \frac{1}{N_{\rm t}} \frac{\partial}{\partial\mu_{\rm q}}\log Z \ , 
 \label{eq:quark_number}
\end{equation}
where $N_{\rm t}$ is the temporal lattice size.
The trace is evaluated by the noise method.
If the excursion problem and the singular drift problem are under control, $\bar{\mathcal{O}}$ is proved to be the expectation value of the original path integral~\cite{Aarts:2009uq,Nishimura:2015pba}.
Otherwise $\bar{\mathcal{O}}$ can converge to a wrong result.
The judging criterion we employ is the drift histogram~\cite{Nagata:2016vkn}, utilizing the magnitude of the drift term~\eqref{eq:drift_term},
\begin{equation}
 v^{(f,g)} := \underset{x,\nu}{\max}
              \sqrt{\frac{1}{3}
              {\rm tr} \, \left( v_{x,\nu}^{(f,g) \dagger} v_{x,\nu}^{(f,g)} \right) } \ .
 \label{eq:drift_magnitude}
\end{equation}
If the magnitude of the drift term falls off exponentially or faster, $\bar{\mathcal{O}}$ is proved to be valid.
The excursion problem is originated from the power-law behavior in the gauge part of the drift, and the singular drift problem in the fermion part.
The former is also measured by the configuration distance from the SU(3) manifold, called the unitarity norm,
\begin{equation}
 \mathcal{N}
 := \frac{1}{12 N_{\rm s}^3 N_{\rm t}}
    \sum_{x, \nu}\mathrm{tr}\, (\mathcal{U}_{x,\nu}^{\dagger} \mathcal{U}_{x, \nu}-\mathbf{1}) \ ,
 \label{eq:unitary_norm}
\end{equation}
where $N_{\rm s}$ is the spatial lattice size.
The excursion problem is reduced by the gauge cooling~\cite{Seiler:2012wz,Aarts:2013uxa},
\begin{equation}
 \delta_g \, \mathcal{U}_{x,\mu} = g_x \mathcal{U}_{x,\mu} g_{x + \hat{\mu}}^{-1}\ , \quad g\in\text{SL(3,C)} \ .
 \label{eq:gauge_cooling}
\end{equation}
It minimizes the unitarity norm.
We employ the gauge cooling at the end of each Langevin update, which does not violate the justification~\cite{Nagata:2015uga,Nagata:2016vkn}.
It has been pointed out that the growth of the unitarity norm does not always lead to the power-law of the gauge drift histogram~\cite{Hirasawa:2020bnl}.
Control of the unitarity norm is a sufficient condition for the correct convergence but is not necessary.

\section{Setup and result}
\label{sec:setup_result}

\begin{figure}
 \centering{}
 \includegraphics[width=7.5cm]{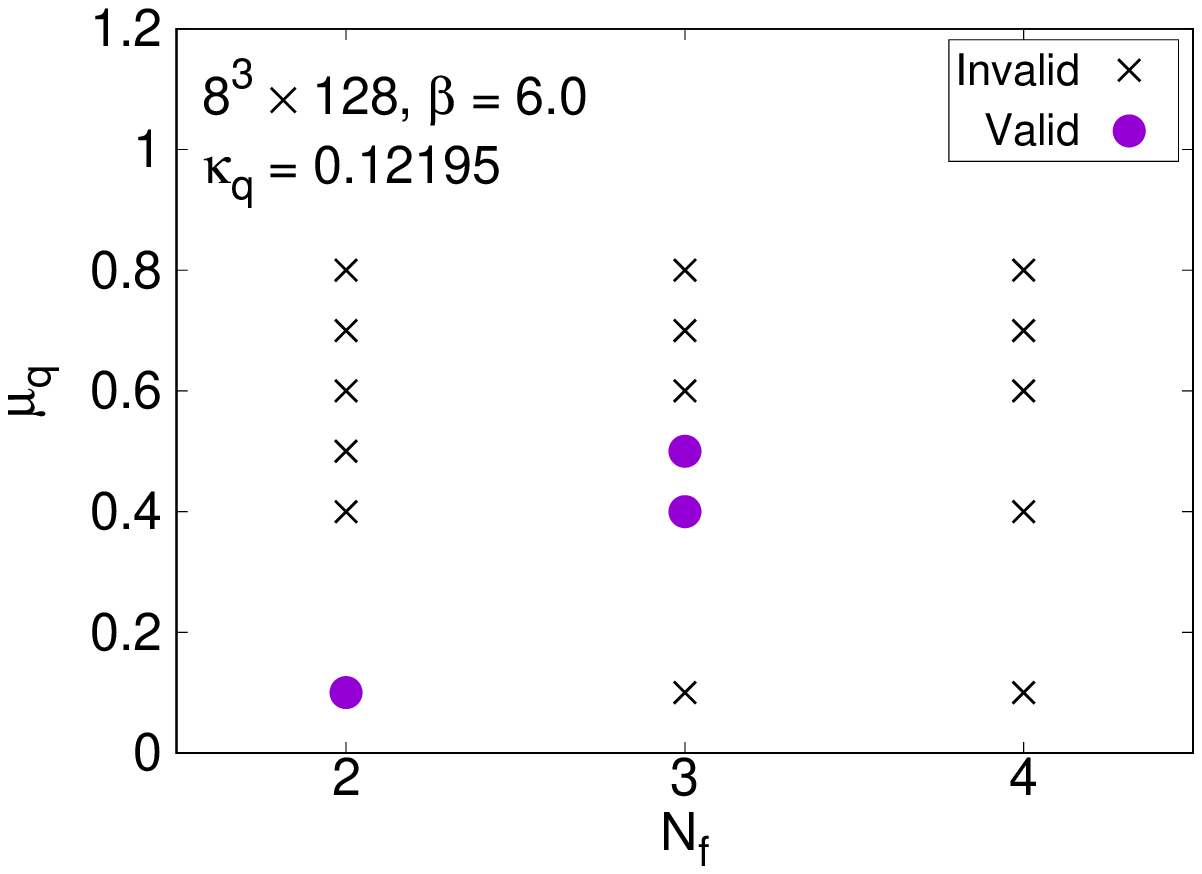}
 \includegraphics[width=7.5cm]{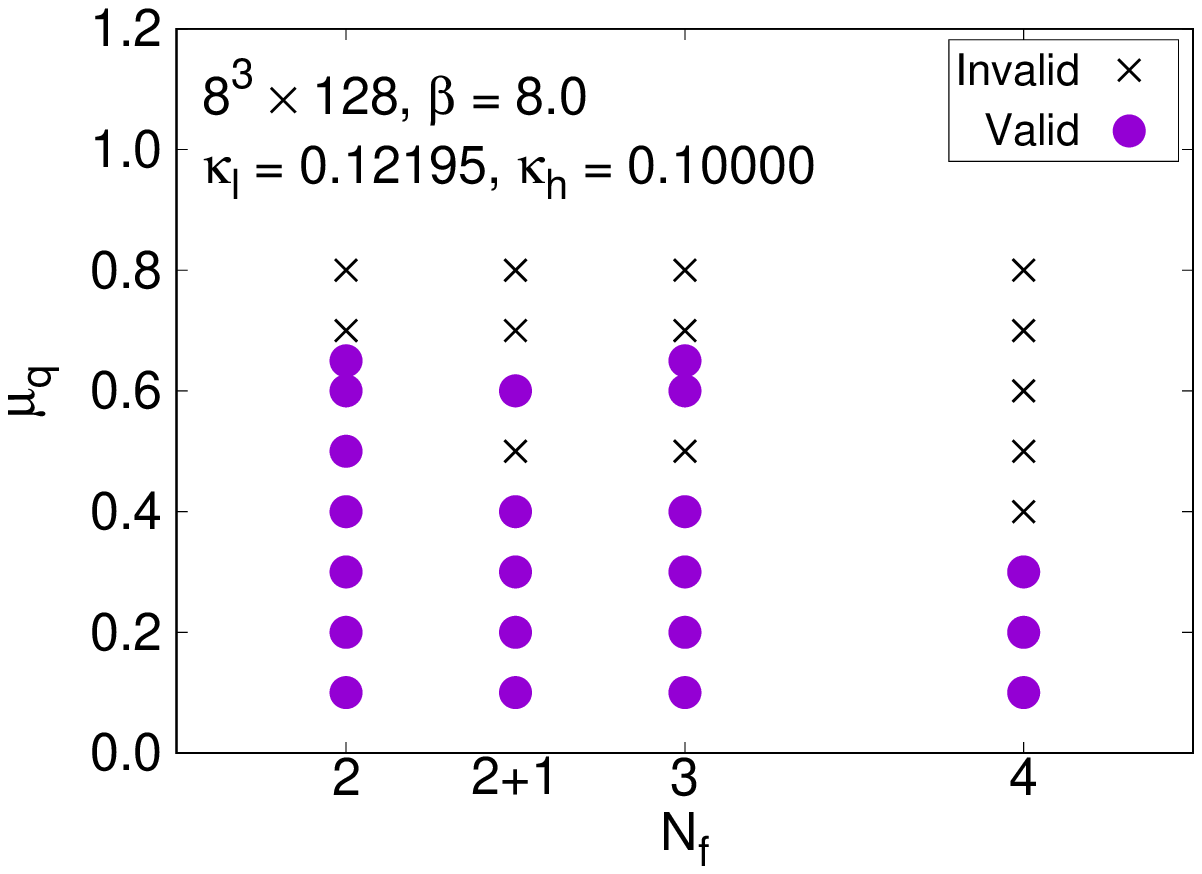}
 \caption{Flavor number $N_{\rm f}$ dependence of the validity region of the CLM at $\beta = 6.0$ (left panel) and $\beta = 8.0$ (right panel) on the $8^3 \times 128$ lattice.
 }
 \label{fig:N_f-validity_region}
\end{figure}

\begin{figure}
 \centering{}
 \includegraphics[width=7.5cm]{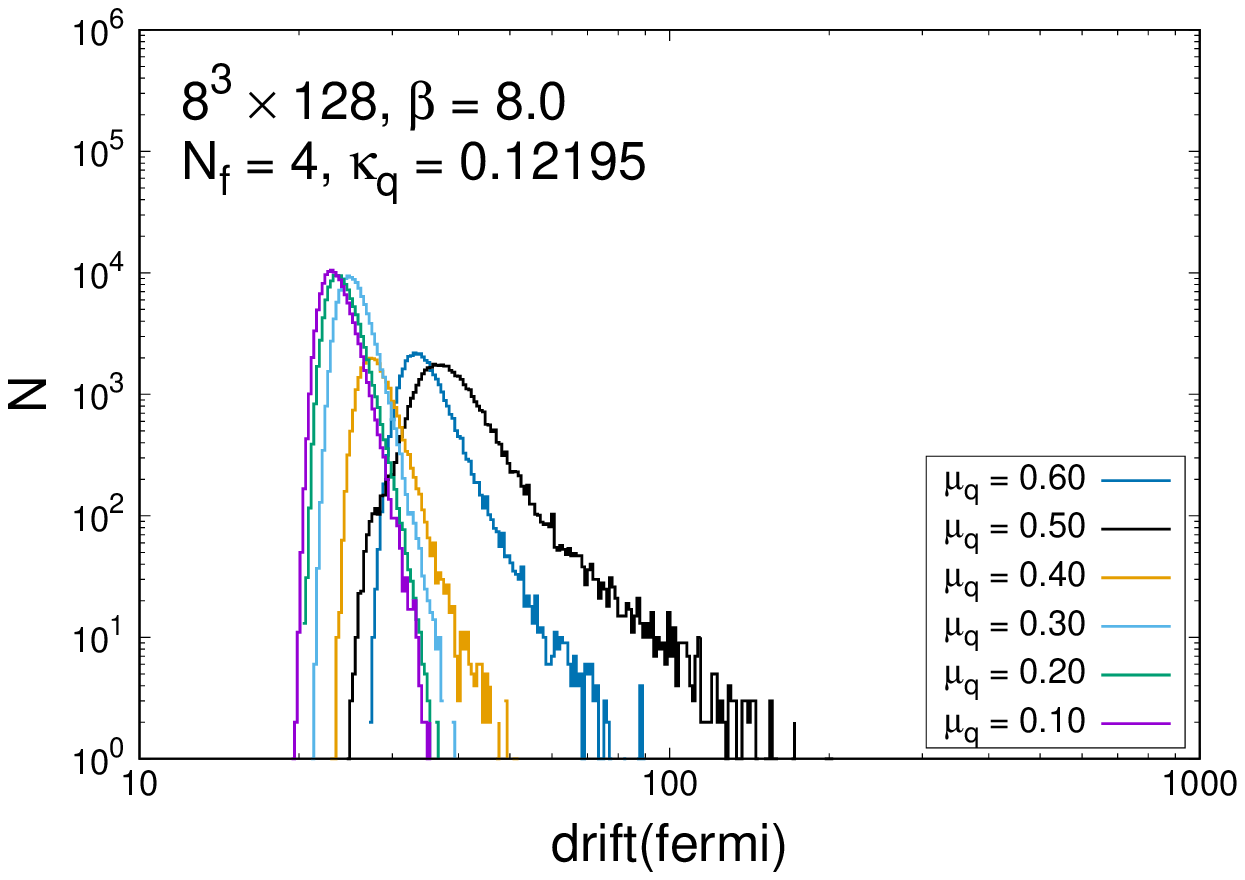}
 \includegraphics[width=7.5cm]{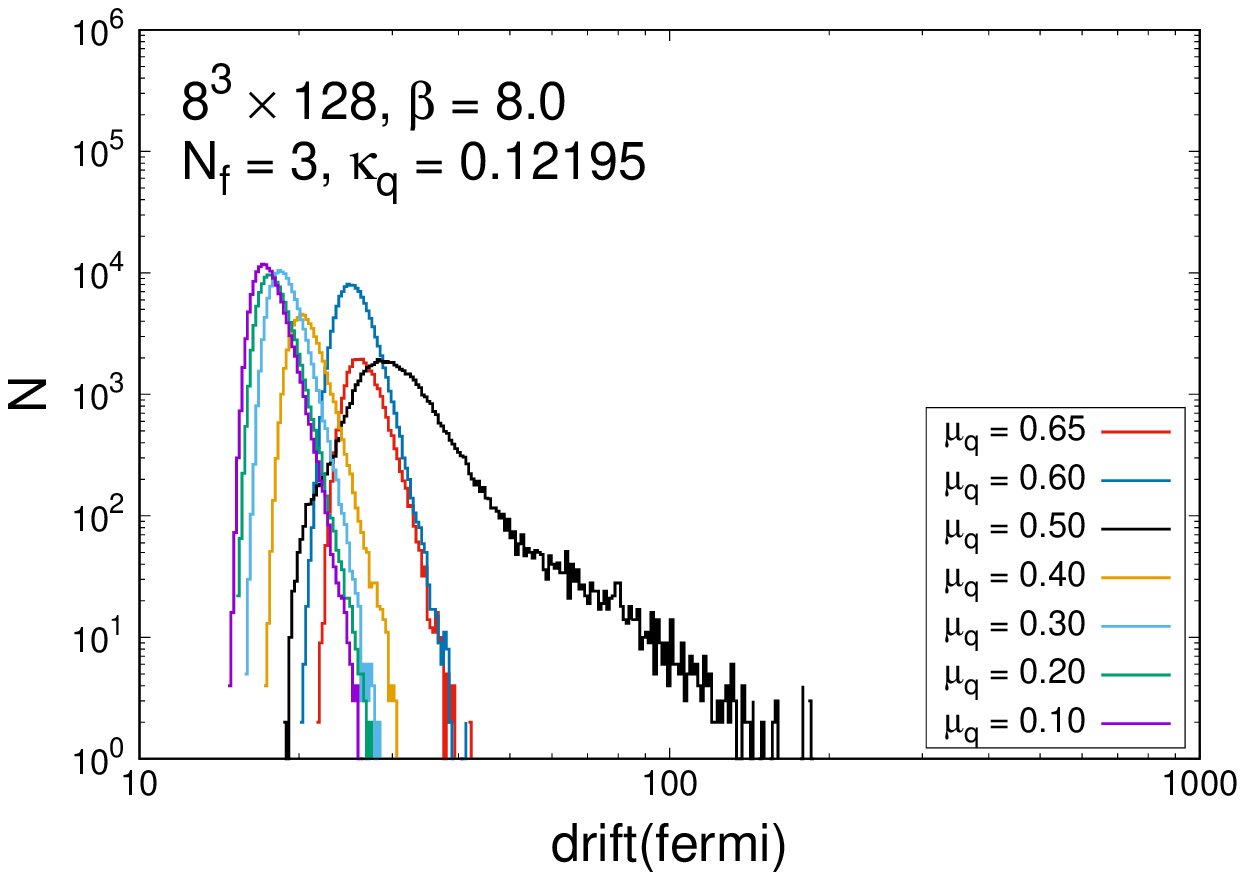}
 \includegraphics[width=7.5cm]{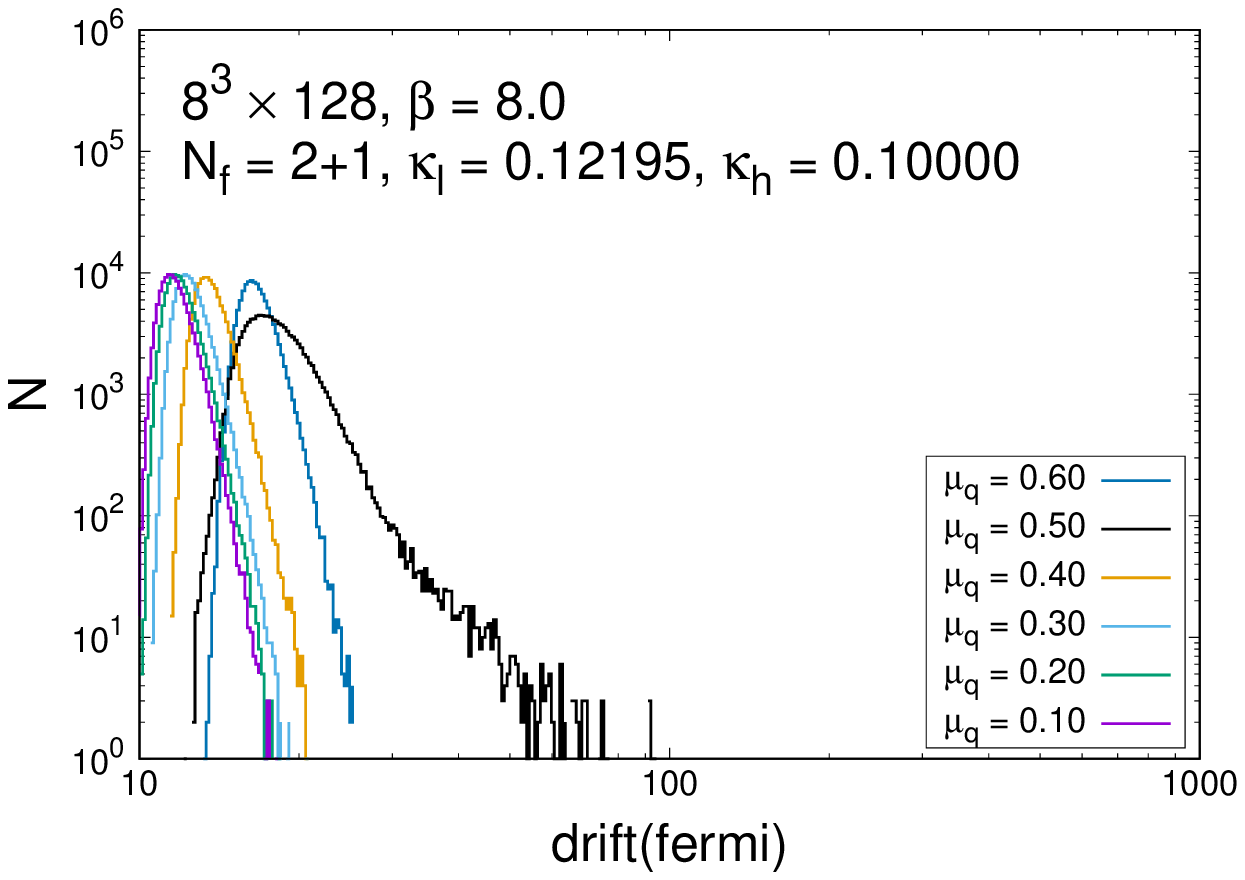}
 \includegraphics[width=7.5cm]{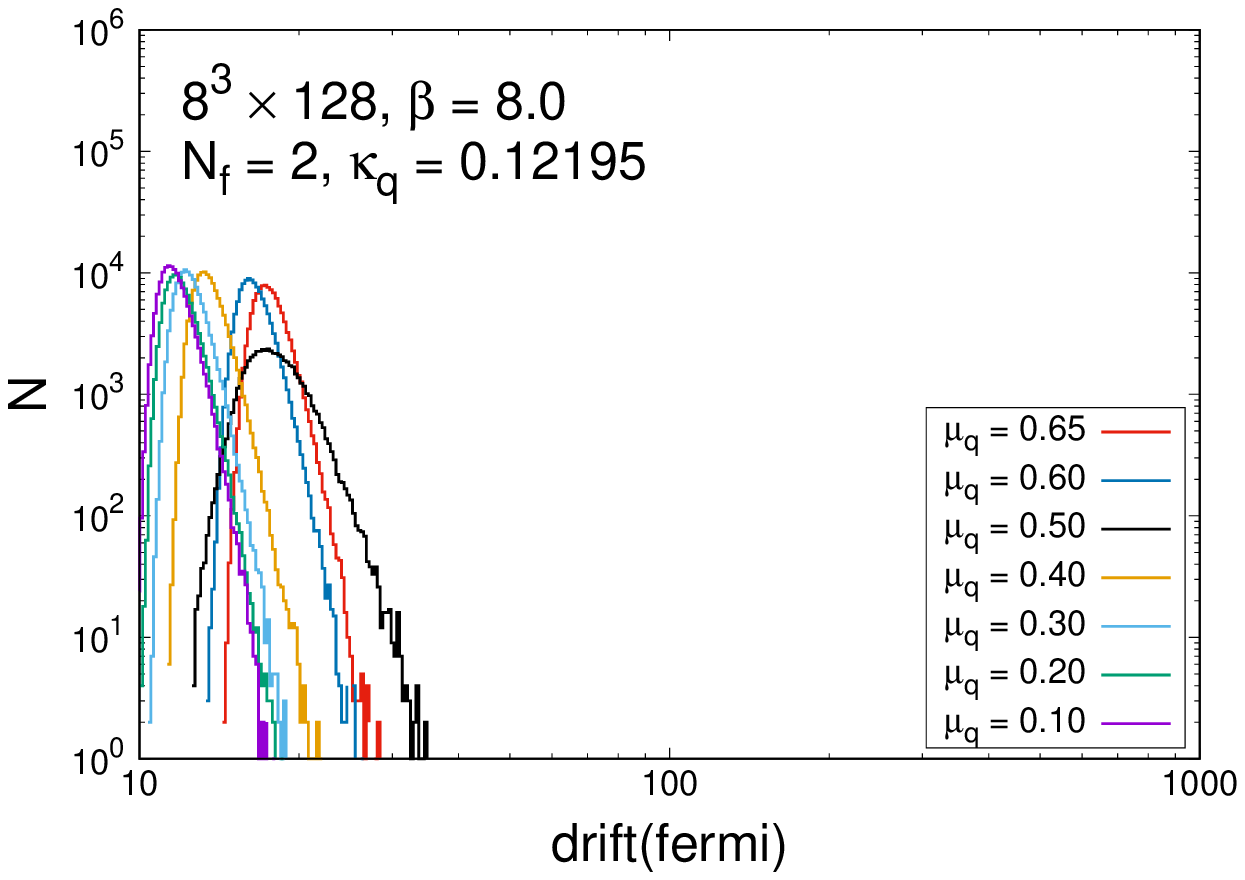}
 \caption{Histograms of the magnitude of the fermion drift term at $\beta = 8.0$ on the $8^3 \times 128$ lattice.
          The flavor number is $N_{\rm f} = 4$ (upper left), $N_{\rm f} = 3$ (upper right), $N_{\rm f} = 2 + 1$ (lower left), and $N_{\rm f} = 2$ (lower right).}
 \label{fig:drift_fermi}
\end{figure}

\begin{figure}
 \centering{}
 \includegraphics[width=7.5cm]{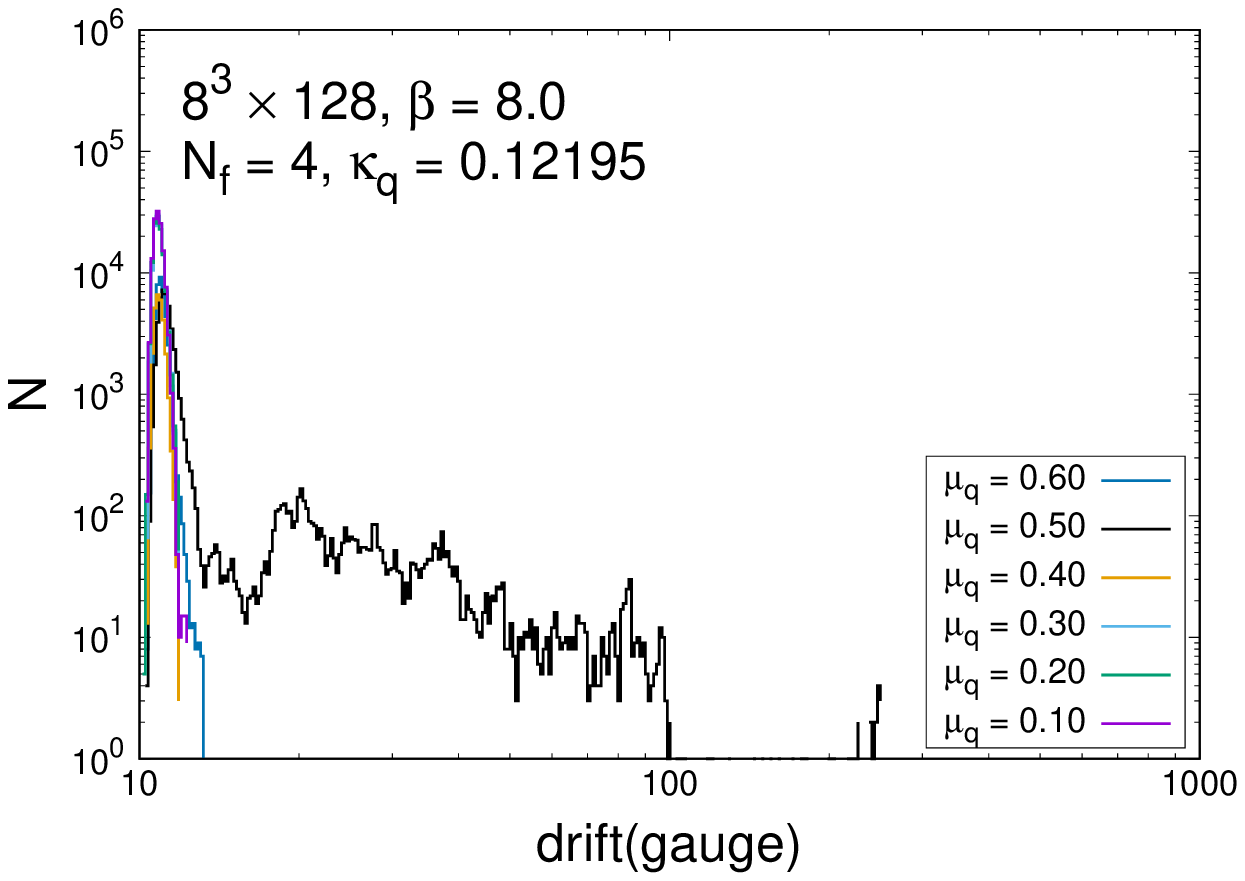}
 \includegraphics[width=7.5cm]{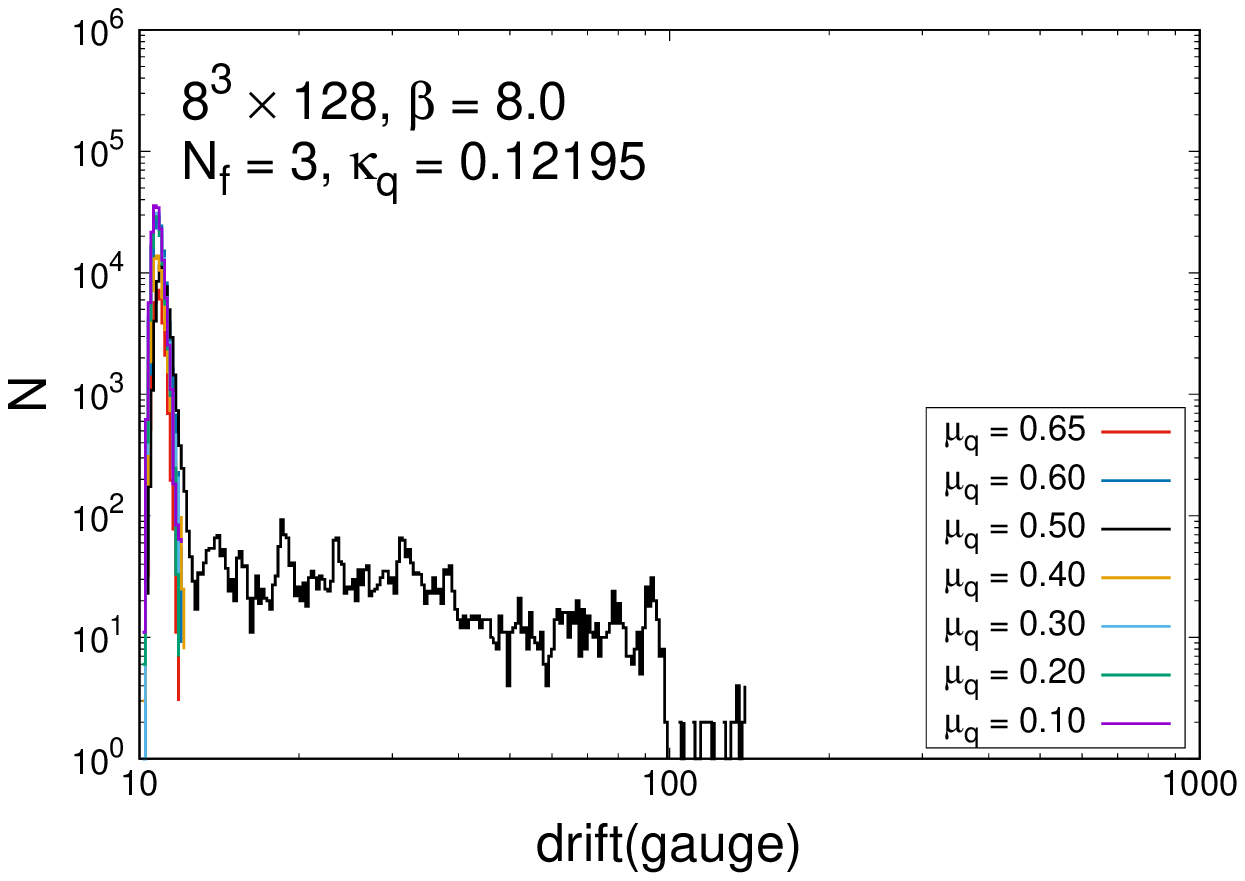}
 \includegraphics[width=7.5cm]{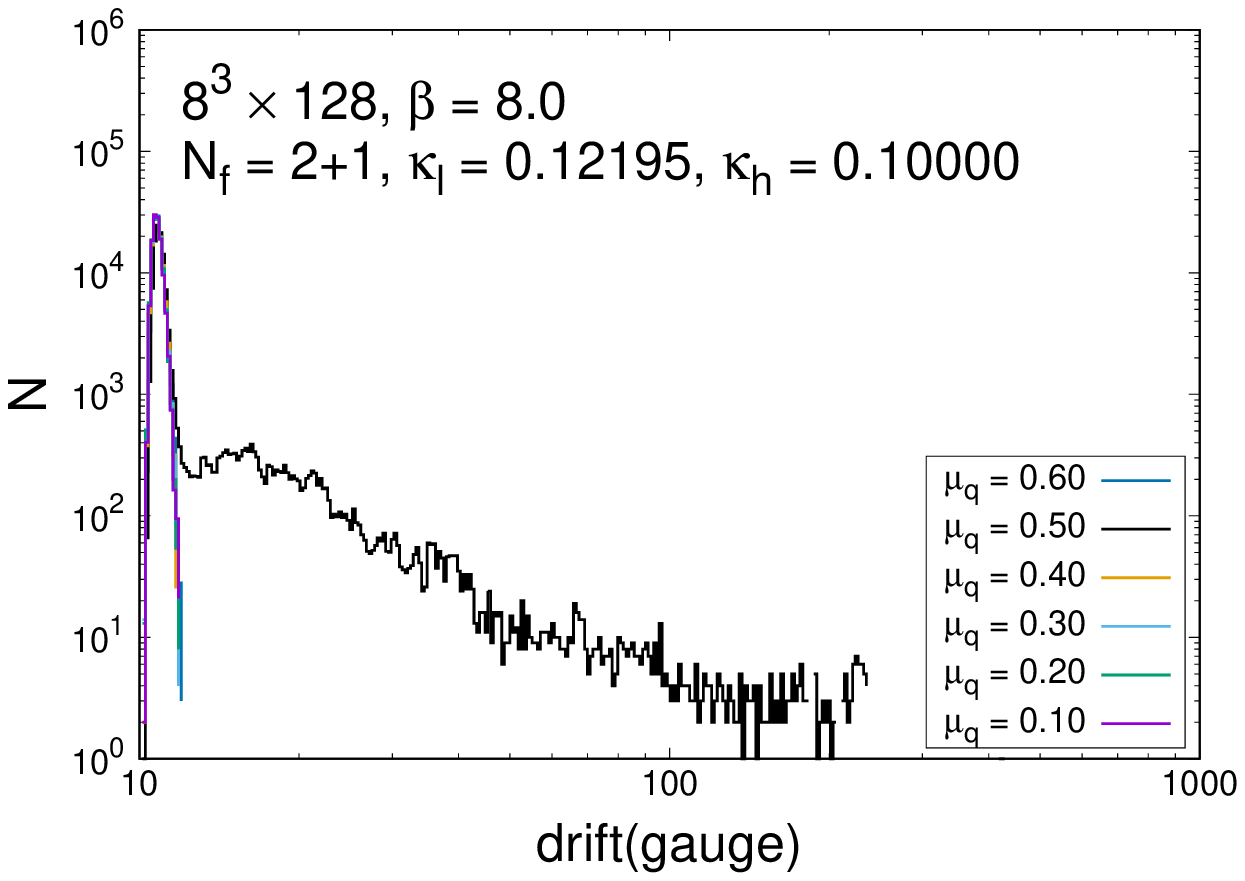}
 \includegraphics[width=7.5cm]{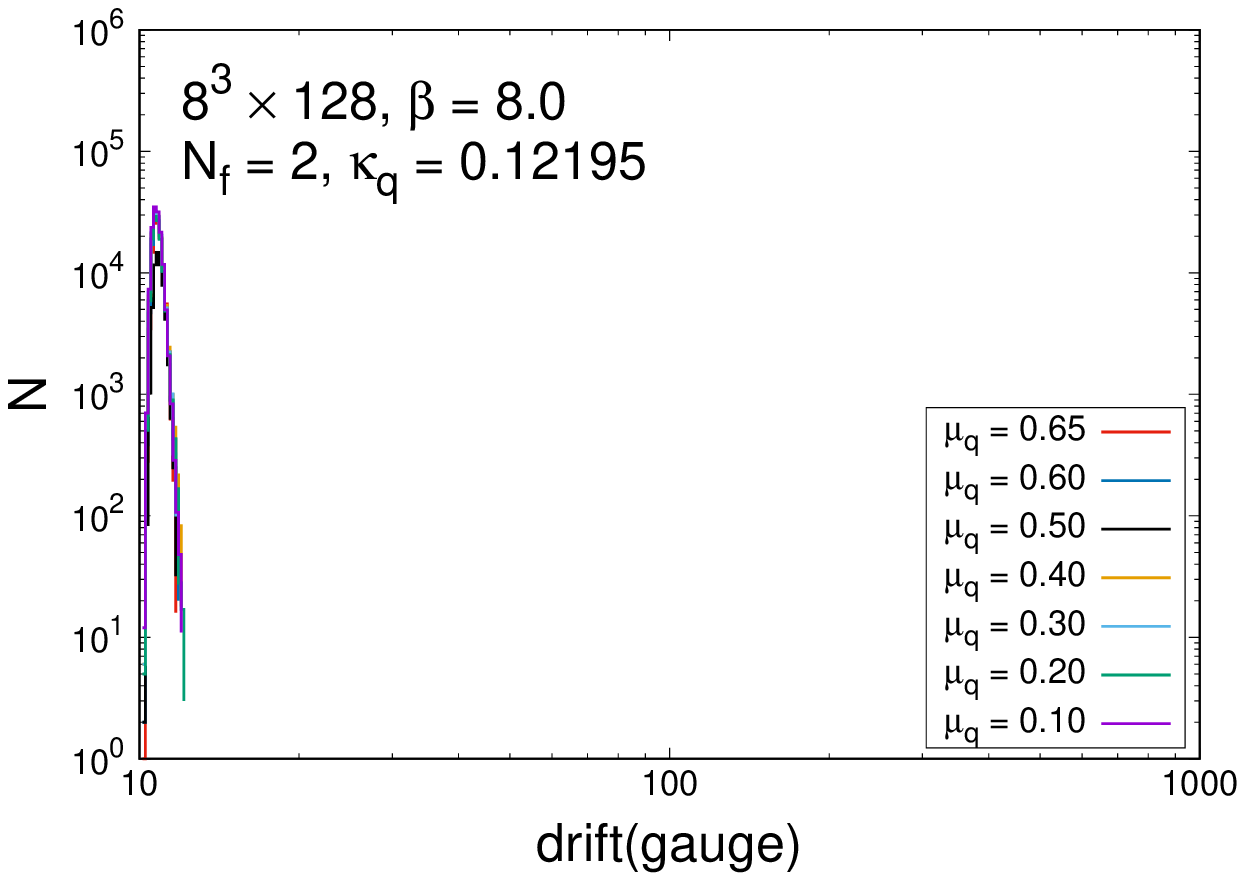}
 \caption{Histograms of the magnitude of the gauge drift term at $\beta = 8.0$ on the $8^3 \times 128$ lattice.
          The flavor number is $N_{\rm f} = 4$ (upper left), $N_{\rm f} = 3$ (upper right), $N_{\rm f} = 2 + 1$ (lower left), and $N_{\rm f} = 2$ (lower right).}
 \label{fig:drift_gauge}
\end{figure}

\begin{figure}
 \centering{}
 \includegraphics[width=7.5cm]{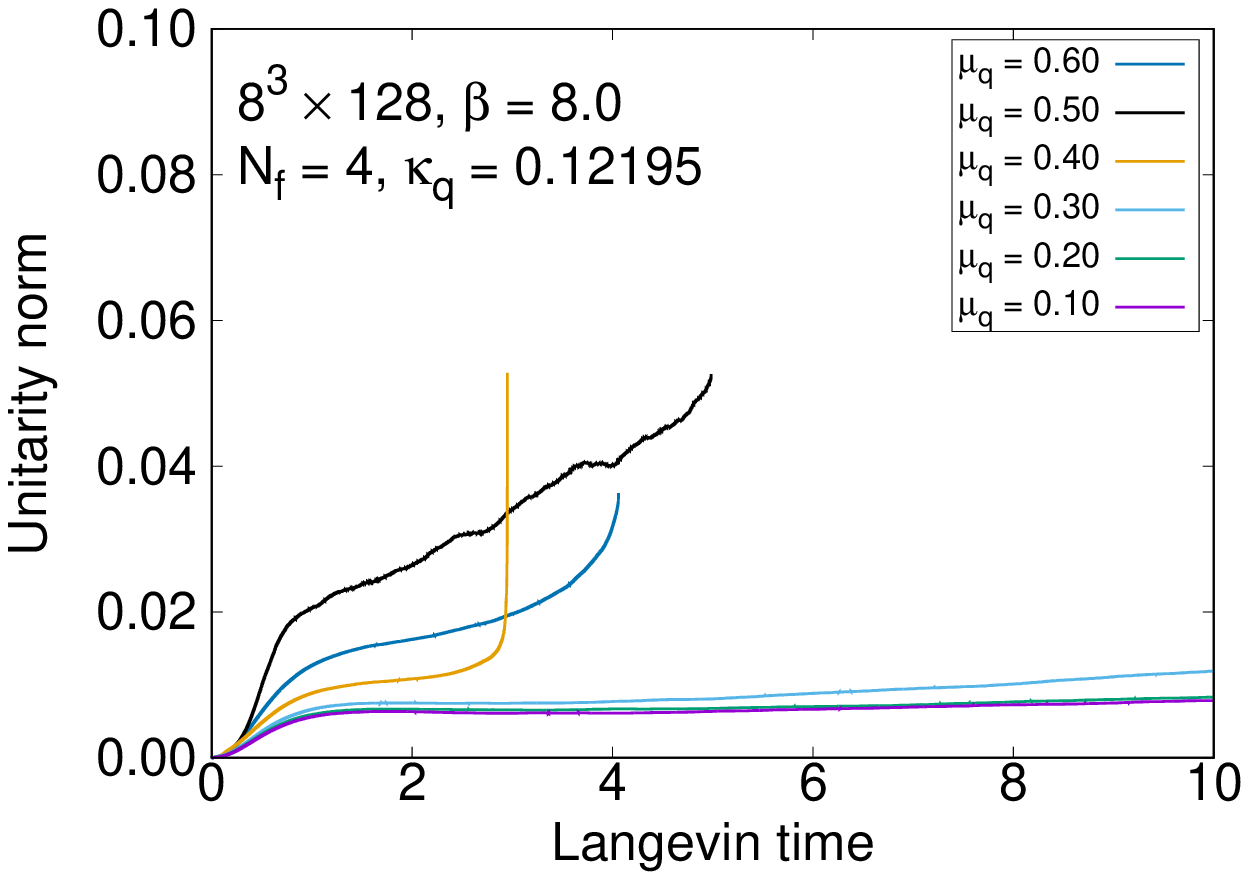}
 \includegraphics[width=7.5cm]{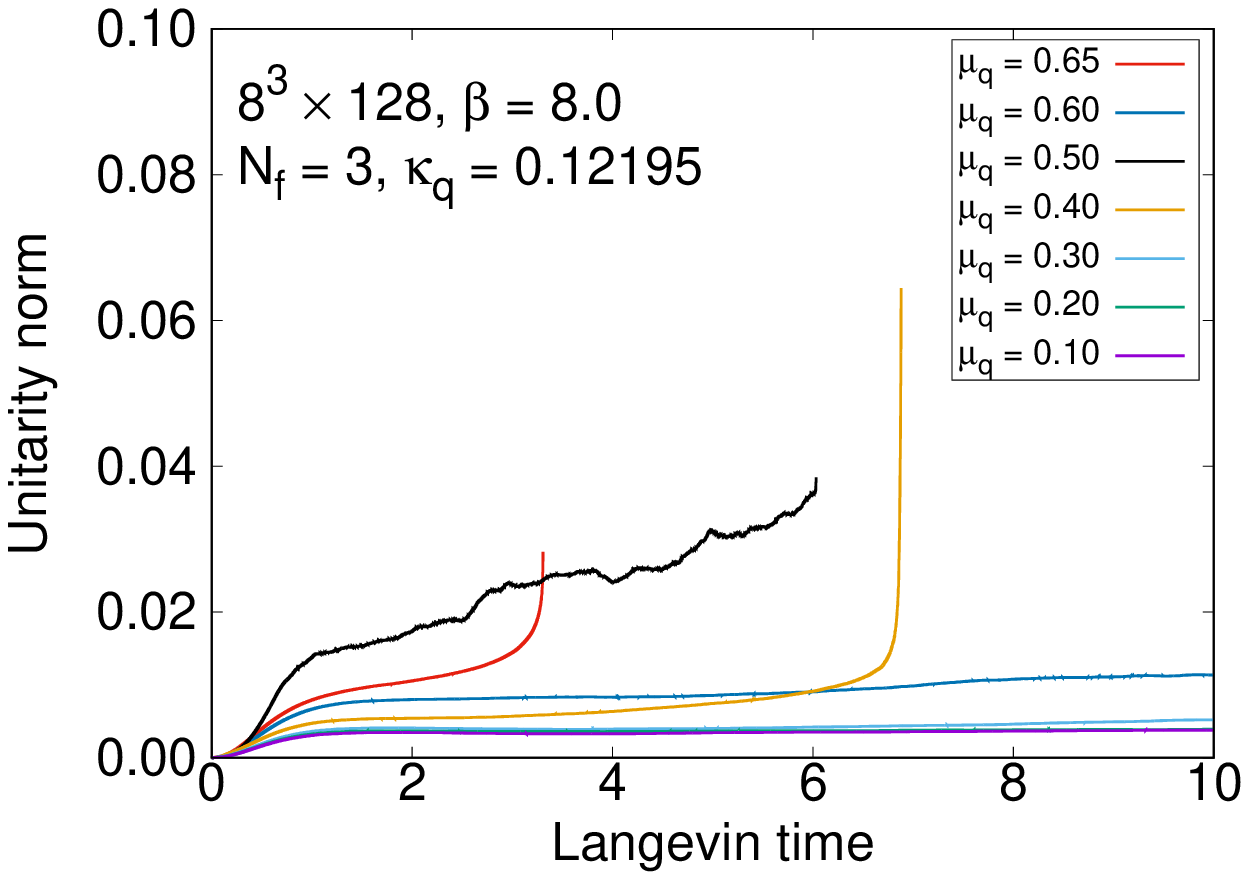}
 \includegraphics[width=7.5cm]{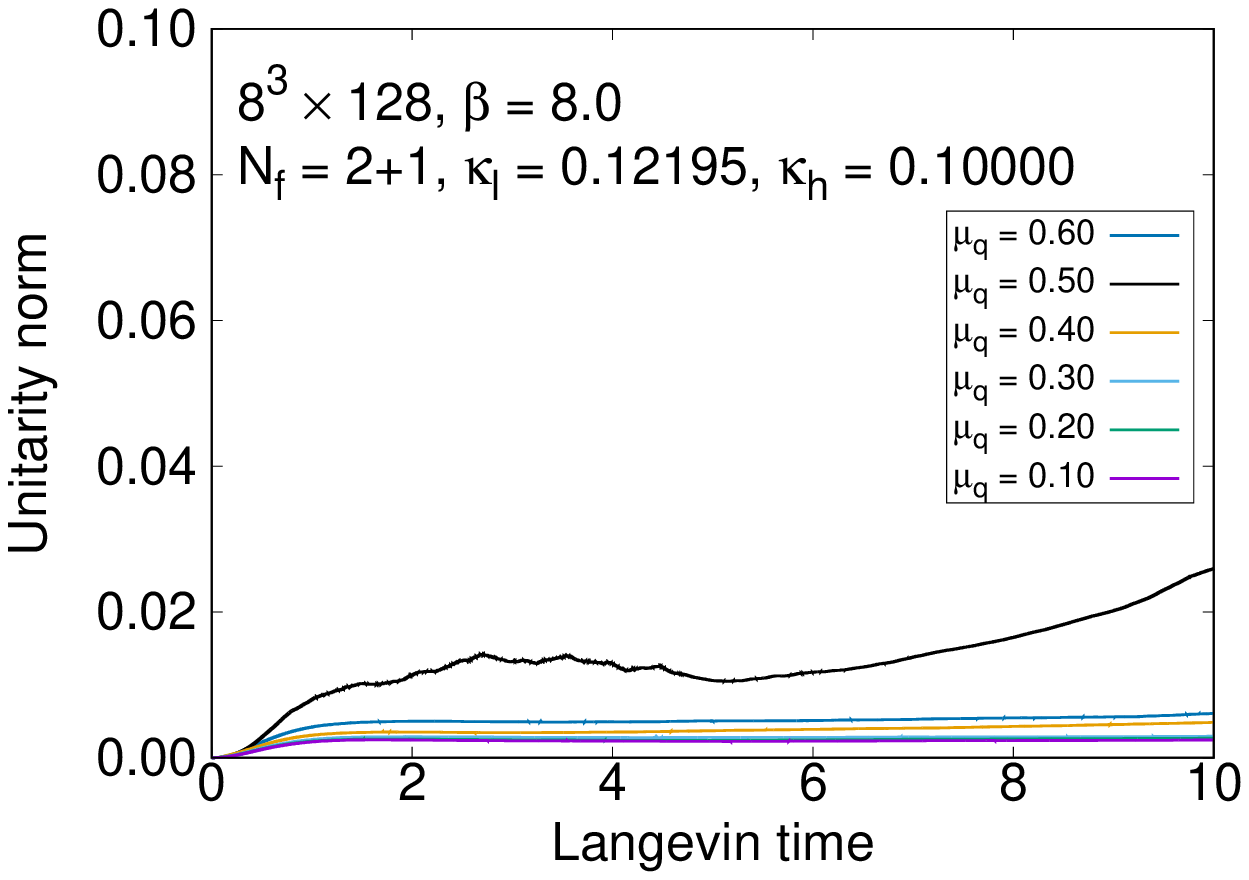}
 \includegraphics[width=7.5cm]{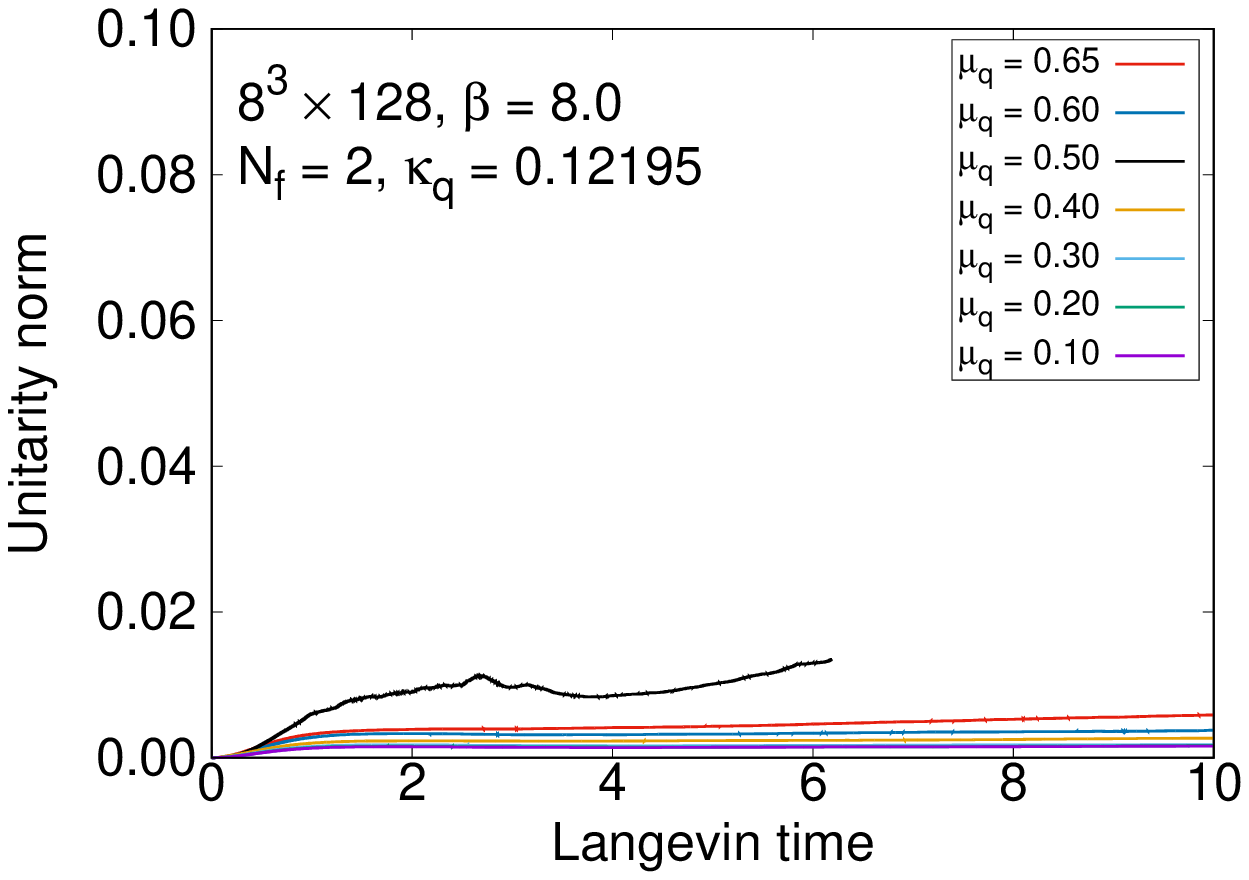}
 \caption{Histories of the unitarity norm~\eqref{eq:unitary_norm} at $\beta = 8.0$ on the $8^3 \times 128$ lattice.
          The flavor number is $N_{\rm f} = 4$ (upper left), $N_{\rm f} = 3$ (upper right), $N_{\rm f} = 2 + 1$ (lower left), and $N_{\rm f} = 2$ (lower right).}
 \label{fig:unitary_norm}
\end{figure}

\begin{figure}
 \centering{}
 \includegraphics[width=9cm]{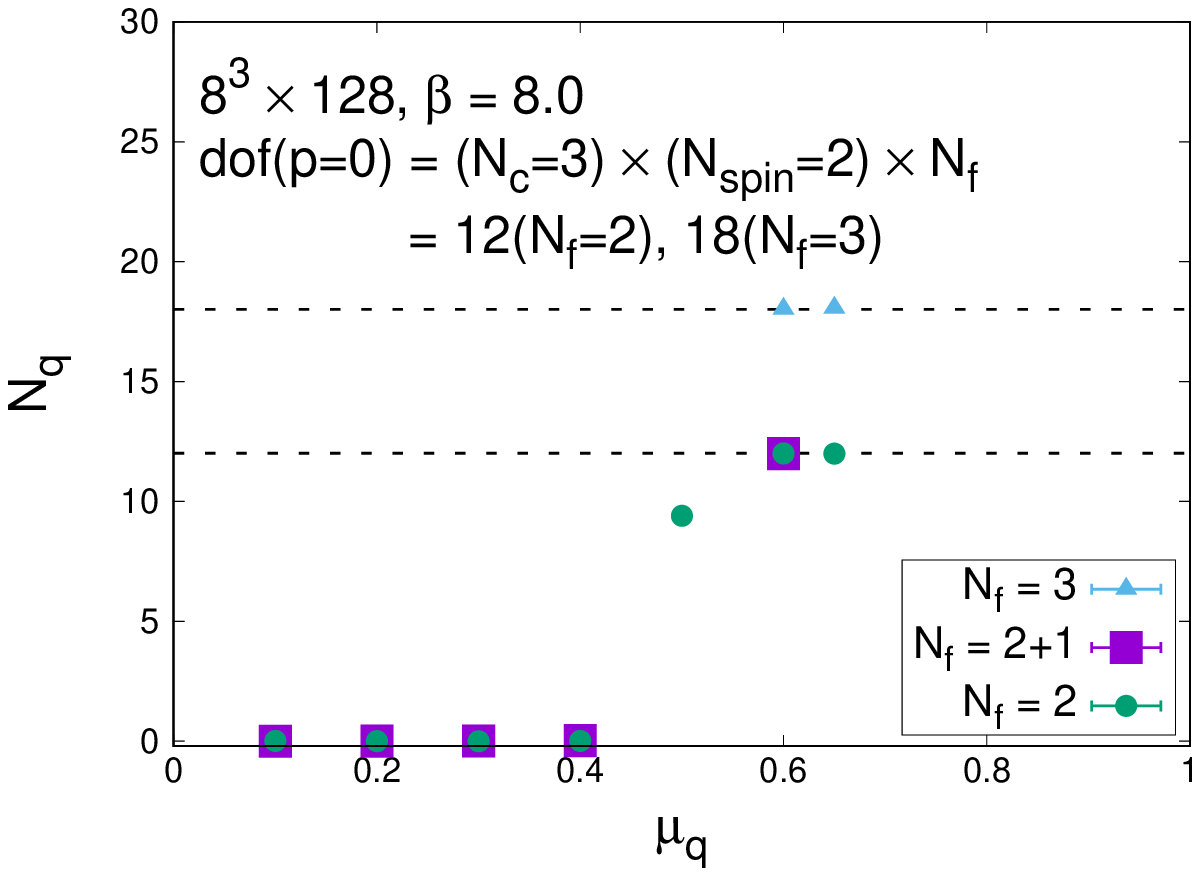}
 \caption{The quark number $N_{\rm q}$ as a function of the quark chemical potential $\mu_{\rm q}$ at $\beta = 8.0$ on the $8^3 \times 128$ lattice.}
 \label{fig:N_q}
\end{figure}

We search for the validity region of the CLM with respect to the flavor number $N_{\rm f} = 2, 2 + 1, 3, 4$, $\beta = 6.0, 8.0$, and $m_{\rm q} = 0.001$--$0.100$ as a function of $\mu_{\rm q}$.
We set $\kappa_c = 1 / 8$ and $m_{s} / m_{ud} = 0.1$ for this exploratory study.
The low temperature is realized by a small aspect ratio of the $8^3 \times 128$ lattice.
The high density corresponds to the large quark chemical potential in our simulation up to $\mu_{\rm q} / T = 102$.
The step size of the complex Langevin evolution is chosen to be $\varepsilon_L = 1.0 \times 10^{-4}$, which is adaptively reduced~\cite{Aarts:2009dg} if the drift magnitude~\eqref{eq:drift_magnitude} is larger than a certain threshold determined at each simulation point.
The total Langevin time is $\tau \leq 10$ after thermalization.
We accumulate data before the rapid growth of the unitarity norm, as explained below.
The error is estimated by the jackknife method with a bin size of $0.4$ Langevin time.

Figure~\ref{fig:N_f-validity_region} represents the flavor number dependence of the validity region of the CLM judged by the drift histogram.
At $\beta = 6.0$, we found small validity regions.
At $\beta = 8.0$, in contrast, we have wider validity regions, as in the case of the previous works~\cite{Kogut:2019qmi,Tsutsui:2021}.
A finer lattice gives better convergence to the correct result by the CLM.
We also found decrease of $N_{\rm f}$ causes enlargement of the validity regions.
It is an encouraging result for the realistic $N_{\rm f} = 2 + 1$ CLM simulation.
We confirmed the validity region is not significantly changed for the quark mass within $m_{\rm q} = 0.001$--$0.100$.
The quark mass dependence is also consistent with the $N_{\rm f} = 4$ staggered fermion result~\cite{Tsutsui:2021}.
It is noticed that the $N_{\rm f}$ dependence in this work does not agree with that of the rooted $N_{\rm f} = 2$ and $4$ staggered fermion at finite temperature on $12^3 \times 6$~\cite{Sinclair:2018rbk}.
The temperature dependence of the validity region shall be evaluated to clarify the origin of the disagreement.
Another remaining task is an analysis along the line of constant physics.
The dependence of the validity region has been still investigated with the fixed bare parameters.
It is also an important task for a definite comparison.

In Figs.~\ref{fig:drift_fermi} and \ref{fig:drift_gauge}, we plot the histograms of the magnitude of the gauge and fermion drift terms~\eqref{eq:drift_magnitude}, respectively.
Data at large chemical potentials are omitted due to the breakdown of the CLM simulation by non-convergence of the iterative solver or floating-point exceptions.
These histograms are used to judge if the CLM result is valid or not.
The histogram of the fermion drift term has a clear power-law behavior at $\mu_{\rm q} = 0.40$--$0.60$ with $N_{\rm f} = 4$.
The CLM result is not guaranteed to be valid at these values of $\mu_{\rm q}$.
On the other hand, the power-law behavior is less observed with $N_{\rm f} = 2, 2 + 1, 3$.
Among these histograms, those at $\mu_{\rm q} = 0.50$ are prominent, where the quark number jumps as presented later.
The power-law is observed in not only the histogram of the fermion drift term but also that of the gauge drift term, except for $N_{\rm f} = 2$.
As in the case of our previous paper~\cite{Ito:2020mys}, the CLM simulation often fails around the transition point of the quark number.

We also display the unitarity norm~\eqref{eq:unitary_norm} in Fig.~\ref{fig:unitary_norm}.
The unitarity norm is under control in the valid region, while it has large values in the invalid region.
Although we use data before the rapid growth of the unitarity norm, the power-law behavior can be still visible in the histogram of the fermion drift term, such as the case at $\mu_{\rm q} = 0.4$ with $N_{\rm f} = 4$.
It is important to examine not only the unitarity norm but also the drift term histogram to confirm the validity of the CLM.

Figure~\ref{fig:N_q} displays the quark number (\ref{eq:quark_number}) at $\beta = 8.0$ on the $8^{3} \times 128$ and lattice.
All the plotted data are in the valid region, satisfying the justification criterion.
As in our previous work~\cite{Ito:2020mys}, the quark number shows plateaus.
The height of the first plateau is consistent with $N_{\rm q} = N_{\rm f} \times N_c \times N_{\rm spin} = 12$ for $N_{\rm f} = 2$, and with $N_{\rm q} = 18$ for $N_{\rm f} = 3$, as expected from the free theory on the lattice.
It is important for a study of the color superconducting phase (CSC) on the lattice, because the formation of the Cooper pair is enhanced on the Fermi surface.
Since the validity region covers the first plateau of the quark number, it suggests the possibility of detecting the CSC non-perturbatively by the CLM.

\section{Summary}
\label{sec:summary}

We explored the validity region of the CLM by varying the flavor number, $\beta$, and quark masses as a function of the quark chemical potential.
The CLM simulation is performed at low temperature and high density QCD by use of $N_{\rm f} = 2, 2 + 1, 3, 4$ Wilson fermions on the $8^{3} \times 128$ lattice.
We confirmed the existence of the validity region, judged by the drift histogram, at low temperature and high density in the range of $\mu_{\rm q} / T = 12.8$--$102$.
It is a crucial first step to unveil the physics of low temperature and high density QCD by the CLM.

The detailed study of $N_{\rm f}$ dependence shows that the validity region is enlarged toward smaller $N_{\rm f}$.
It is an encouraging result that realistic $N_{\rm f} = 2 + 1$ QCD has a broader validity region than that of $N_{\rm f} = 4$ case, which has been extensively studied by the staggered fermion.
Our $N_{\rm f}$ dependence is opposite to the case of the rooted $N_{\rm f} = 2$ and $4$ staggered fermion at finite temperature~\cite{Kogut:2019qmi}.
The difference is probably originated from the finite temperature effect.

We also found the validity region is enlarged by the change of $\beta = 6.0$ to $\beta = 8.0$ with the bare parameters being fixed.
It is consistent with the tendency in the $N_{\rm f} = 4$ staggered fermion~\cite{Kogut:2019qmi,Tsutsui:2021}.
A finer lattice leads to better control of the validity.
The result suggests that the action improvement may lead to wider validity regions.

We measured the quark number in the validity region, and confirmed the validity region reaches the first plateau, as in our previous study with $N_{\rm f} = 4$ staggered quark~\cite{Ito:2020mys}.
The height of the first plateau is found to be proportional to $N_{\rm f}$, as expected from the lattice free theory.
This confirmation is relevant for a study of the CSC on the lattice since the Cooper pair forms easily on the Fermi surface.
The result of this work supports the possibility of a non-perturbative study of the CSC in the validity region of the CLM.

An important future work is to increase the spatial lattice volume, which is required to fully extract the non-perturbative effect.
We still need to check the spatial volume dependence of the validity region.
Another direction is the study of CSC by the lattice simulation.
Our perturbative study predicts the critical coupling for CSC on the lattice~\cite{Yokota:2021lat}.
Non-perturbative confirmation along this line is ongoing by the CLM with the staggered fermion~\cite{Tsutsui:2021lat}.
Since the flavor number plays an important role in CSC, usage of the Wilson fermion is advantageous, which has an exact flavor symmetry.
We hope to clarify the structure of CSC non-perturbatively.

\acknowledgments
This work was supported by JSPS KAKENHI Grant Numbers JP16H03988, JP21K03553.
S.\ T.\ and T.\ Y.\ were supported by the RIKEN Special Postdoctoral Researchers Program.
Numerical computation was carried out on the Oakbridge-CX provided by the Information Technology Center at the University of Tokyo through the HPCI System Research project (Project ID: hp200079, hp210078) and the Yukawa Institute Computer Facility.


\bibliographystyle{JHEP}
\bibliography{ref}

\end{document}